\documentclass[epj]{svjour}
\usepackage{graphicx}   
\usepackage{dcolumn}    
\usepackage{xcolor}     
\usepackage{epstopdf}   
\usepackage{mathtools}  
\usepackage{url}        
\graphicspath{{graphics/}}       
\usepackage{hyperref}   
\usepackage{booktabs}   
\usepackage{tabularx}   
\usepackage{multirow}   %
\usepackage[caption=false]{subfig} 
\usepackage{cite}

\usepackage[title,titletoc]{appendix} 
\newcommand{\keV}{{\rm \,keV}}

\begin{document}
\title{Influence of NaI background and mass on testing the DAMA modulation}
\author{Madeleine J. Zurowski\inst{1,2}\thanks{madeleine.zurowski@unimelb.edu.au} \and Elisabetta Barberio\inst{1,2}}        
\institute{School of Physics, University of Melbourne, Victoria, Australia \and ARC Centre of Excellence for Dark Matter Particle Physics}
\date{Received: date / Revised version: date}
%
\abstract{
We present here the model dependent and independent sensitivity studies for NaI detectors designed to test the DAMA result, and compare the predicted limits from SABRE with the present performance of both ANAIS and COSINE. We find that the strongest discovery and exclusion limits are set by a detector with the lowest background (assuming equal live times), and also note that our method correctly computes the present exclusion limits previously published by ANAIS and COSINE. In particular, with a target mass of 50 kg and background rate of 0.36 cpd/kg/keV (after veto), SABRE will be able to exclude the DAMA signal with 3$\sigma$ confidence or `discover' it with 5$\sigma$ confidence within 2 years. This strongly motivates the quest for ever lower backgrounds in NaI detectors.
\PACS{
      {PACS-key}{describing text of that key}
     } 
} 
\maketitle


\section{Introduction}
\label{sec:intro}
There is a plethora of evidence that suggests some long-lived, non-luminous, particulate matter within the Universe. These cosmological observations exist on a wide range of scales, and there is no member of the Standard Model of Particle Physics (SM) able to successfully explain them all. As such, a new component called Dark Matter (DM) is required. Of the possible additions to the SM, the Weakly Interacting Massive Particle (WIMP) is one that has enjoyed popularity since its conception. This is due to the fact that particles on the weak scale have a self annihilation rate that naturally gives rise to the observed present day abundance of DM. Despite the fact that the standard assumptions for a WIMP (mass $\sim$ GeV/c$^2$-TeV/c$^2$, cross section $\sim 10^{-40}$ cm$^2$) are now very heavily constrained experimentally, there are still a large number of experiments conducting searches for them.\\
WIMPs are typically probed for in one of three ways; indirect detection (observation of DM annihilation into SM particles), collider searches (observation of DM produced via SM annihilation), or direct detection (scattering of DM off some SM target). In particular, DM direct detection has a distinctive `smoking gun' - a modulating interaction rate due to the relative velocity of the Earth and DM halo as the former rotates about the Sun. This modulation should be observed regardless of the particle interaction model considered as it depends only on the DM velocity distribution. Such a modulation should have a period of a year, and an amplitude on the order of a few percent of the constant rate.\\
One experiment has consistently observed such a signal over 14 annual cycles. Using a NaI target, the DAMA Collaboration has reported evidence of a DM consistent modulation with a combined statistical significance of 12.9$\sigma$ after around 15 years of data taking \cite{Bernabei2018}. Despite this seemingly strong evidence, these results are in increasing tension with other direct detection experiments, which continue to publish null results for DM models that fit the DAMA data \cite{Tanabashi:2018oca}. This has strongly constrained the standard WIMP assumptions, and pushed to more and more particular tailored models to explain these null results consistently.\\ 
The experiments that constrain the DAMA result depend largely on detectors that use a different scattering target, and as such, require the a priori assumption of some particular interaction model for analysis. In order to conclusively dismiss the DAMA signal as having a DM origin, it must be assessed in a model independent manner; that is, through the use of another NaI target experiment attempting to observe this annual modulation. At present, three room temperature detectors are planned (SABRE \cite{Antonello_2019}) or already in the data taking stage (ANAIS and COSINE \cite{amare2021,adhikari2021}), as well as a cryogenic detector (COSINUS \cite{Reindl_2020}) able to distinguish between electron and nuclear recoils being planned. As these all use the same target as DAMA, under the assumption that all the experiments have a threshold efficiency and detector resolution similar to that of DAMA, they are expected to be able to observe the same modulation signal, given a low enough background and long enough data taking period. Thus, model independent sensitivity limits can be constructed in the background-live-time (live-time being the period over which the detector is actively taking data) plane similar to how they are in the $m_{\chi}$-$\sigma_p$ plane in the model dependent case. This allows for a comparison of the performance of the three NaI detectors given their different background and exposure masses.\\
This paper commences by detailing the method used to carry out both model dependent and independent sensitivity calculations in Section \ref{sec:sensecalc}. It then presents the results of both sensitivity cases, and compares the exclusion and discovery powers of the room temperature ANAIS, COSINE, and SABRE in Section \ref{sec:results}, before concluding in Section \ref{sec:conclusions}.

\section{Sensitivity methodology}
\label{sec:sensecalc}
\begin{figure}[!h]
    \centering
    \includegraphics[width=0.5\textwidth]{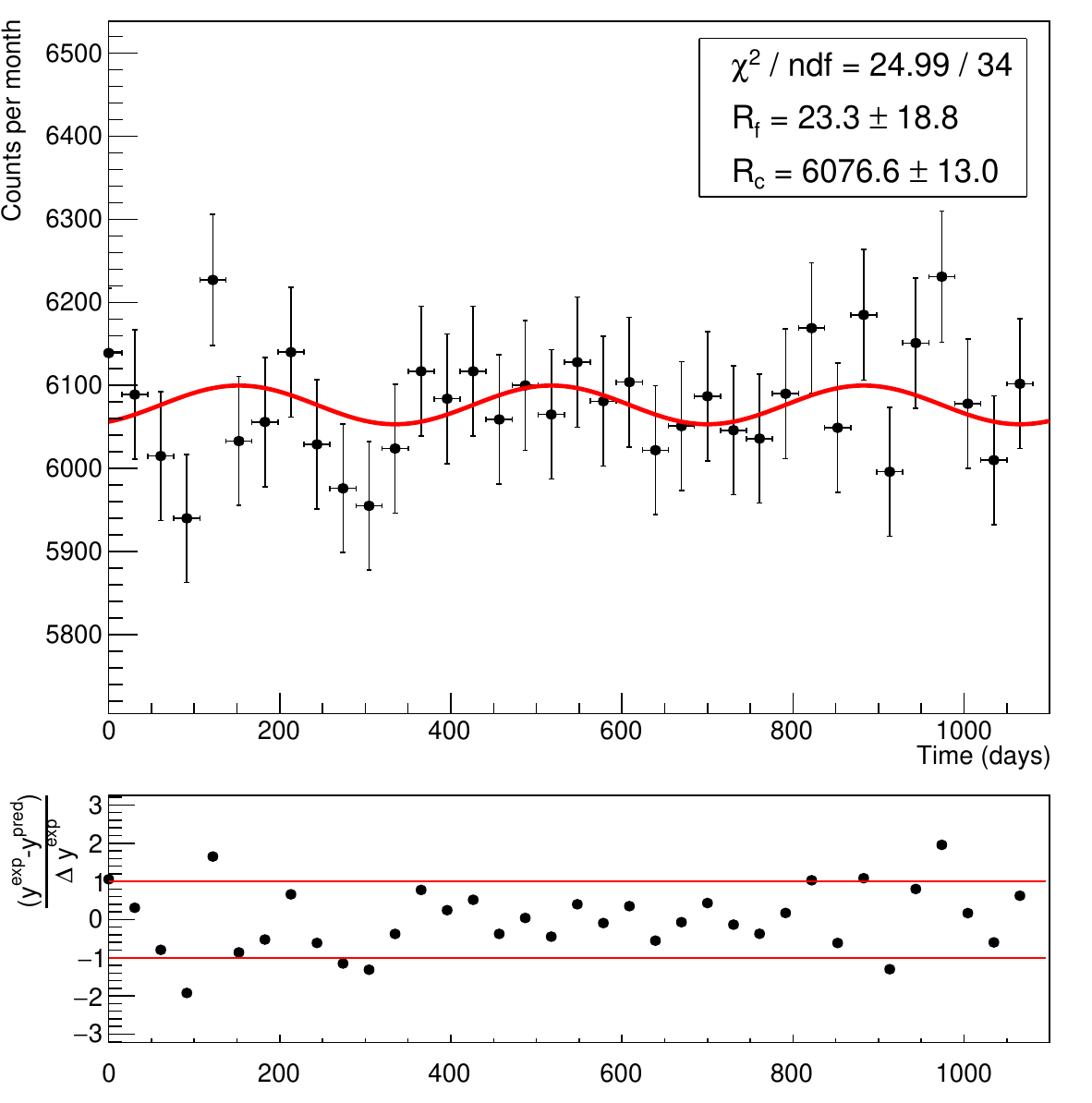}
    \caption{Simulated fluctuations in background only model, with $R_b=1$ cpd/kg/keV over 3 years fit to $R(t)=R_c+R_f\cos{\omega t}$.}
    \label{fig:sample_back_fluctuation}
\end{figure}

\begin{figure}[!h]
    \centering
    \includegraphics[width=0.5\textwidth]{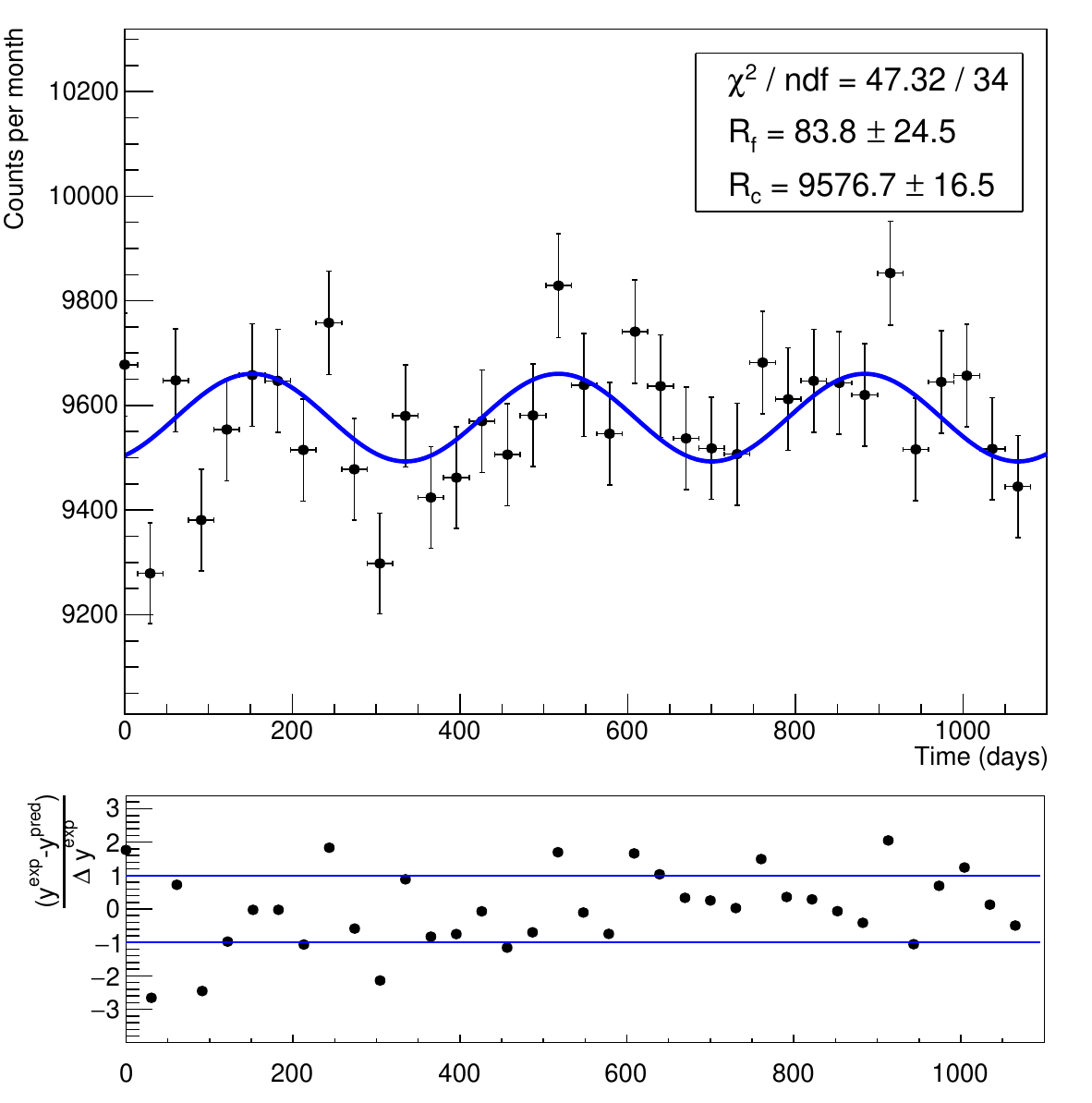}
    \caption{Simulated fluctuations in signal+background model, with $R_b=1$ cpd/kg/keV, $R_m=0.012$ cpd/kg/keV, and $R_0=0.6$ cpd/kg/keV.}
    \label{fig:sample_sigback_fluctuation}
\end{figure}

In general, an experiment will be sensitive to a DM modulating signal if the signature modulation is observable over the detectors background. In this case, the apparent modulation of the background signal due to statistical fluctuations is more problematic than the background rate itself, as this can interfere with the fitting of a modulation component. As such, to assess the sensitivity of a detector, these fluctuations should be modelled over its live time, and fit to a cosine function to find the probability of this masking or mimicking the DM signal. \\
Assuming that in some energy bin of width $\Delta E$ over a time bin period of $\Delta T$ (we assume 30 days unless otherwise stated), and a background rate given by $R_b$ (in units of cpd/kg/keV), these background statistical fluctuations can be modelled by sampling from a Poisson distribution centred on 
\begin{equation}
    N_b = M_E \times \Delta T \times \Delta E \times \epsilon\times R_b,
    \label{back_n}
\end{equation}
where $M_E$ is the mass of the detector in question, and $\epsilon$ its efficiency. The signal plus background (s + b) modulation can be modelled in a similar way, sampling instead from
\begin{equation}
    N_{s+b} = M_E \times \Delta T \times \Delta E \times \epsilon\times\left(R_b+ R_0 +R_m\cos{\omega t} \right),
    \label{sigback_n}
\end{equation}
where $R_0$ and $R_m$ are the constant and modulating signal events observed in a given energy bin. Generating this for every live bin period of detector operation, the total observed rate is then fit to $R(t)=R_c+R_f\cos{\omega t}$ in order to separate the constant ($R_c$) and fluctuating ($R_f$) components. A non-zero modulation can be seen in both the background only and signal plus background cases - see, for example, Figs. \ref{fig:sample_back_fluctuation} and \ref{fig:sample_sigback_fluctuation}, where a 50 kg detector running for 3 years observes a signal compatible with $R_f=23.3 \pm 18.8$ counts per month for background only, and $R_f=83.8 \pm 24.5$ counts per month for signal plus background. 

\begin{figure}[!h]
    \centering
    \subfloat[\label{back_pdf}]{%
        \includegraphics[width=0.5\textwidth]{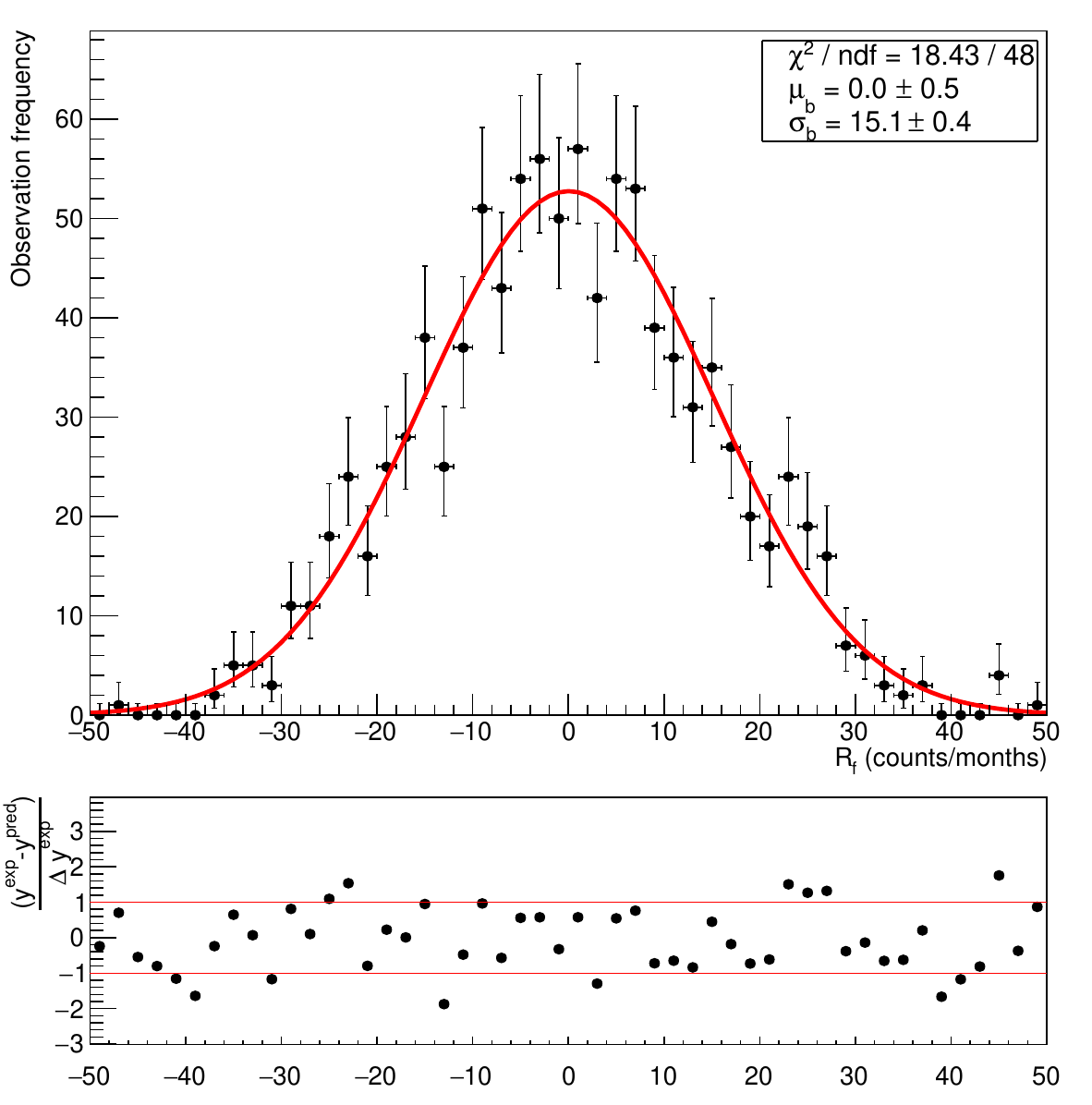}}\\
    \subfloat[\label{sb_pdf}]{%
        \includegraphics[width=0.5\textwidth]{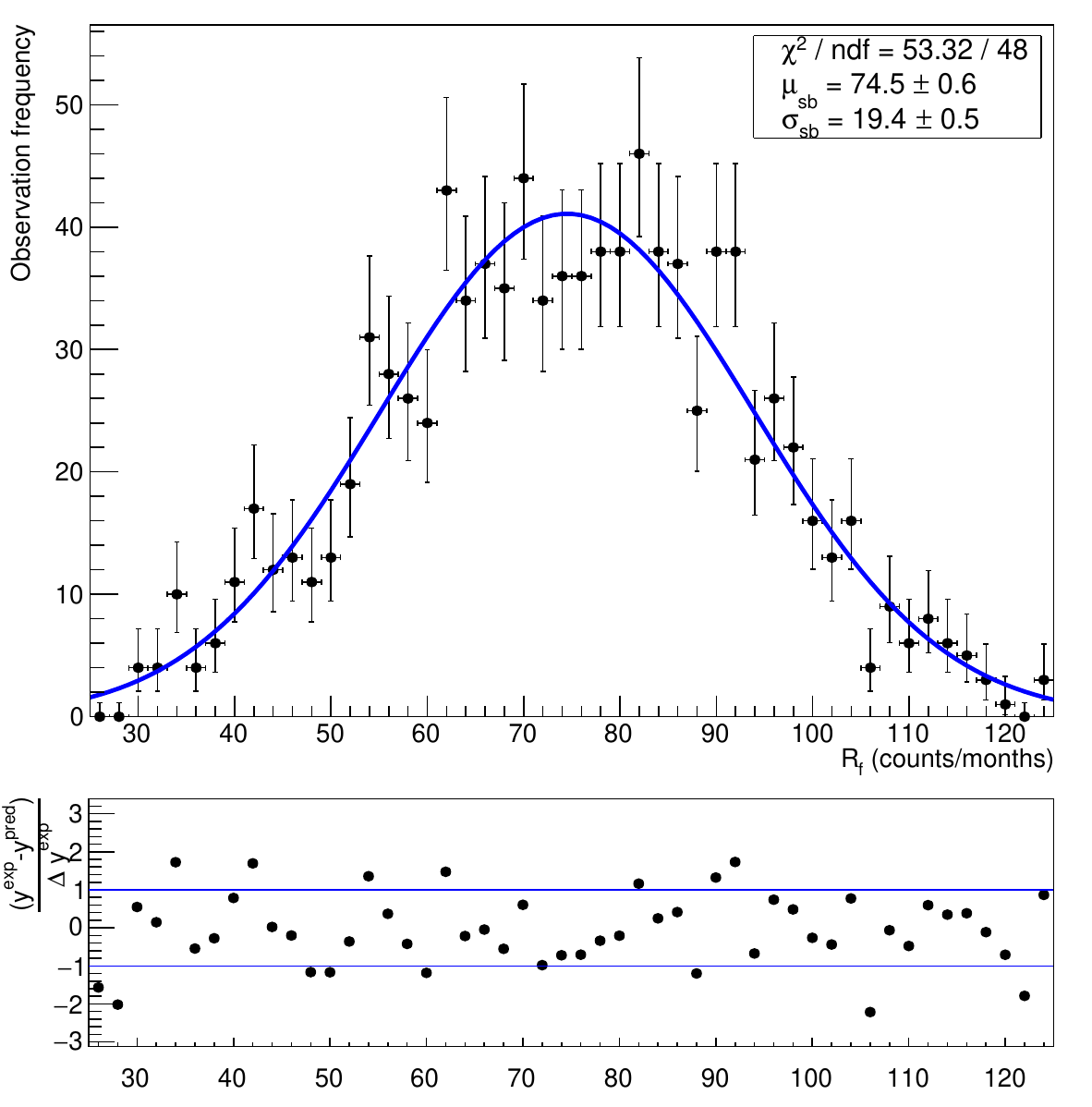}}
    \caption{Background only (top) and signal+background (bottom) distributions, for $R_b=1$ cpd/kg/keV over 3 years, and signal given by DAMA. These are plotted as a function of the number of modulation events observed per month.}
    \label{fig:model_indep_method}
\end{figure}

This modulation over the live time of the detector is then generated a number of times ($\mathcal{O}(100-1000)$), saved, and fit to a gaussian given by 
\begin{equation}
    f(R_f)\propto\exp{\left[-\frac{(R_f-\mu_{(s)b})^2}{2\sigma_{(s)b}^2}\right]}.
\end{equation}
This gives a probability distribution function for the observed modulation amplitude $R_f$ for either the background only ($b$) or signal+background ($sb$) case. Samples of these can be seen in Fig. \ref{fig:model_indep_method}, where we have taken $R_m=0.012$ cpd/kg/keV, $R_0=0.6$ cpd/kg/keV and $\epsilon=1$.
These distributions can then be compared to assess the ability of the detector to distinguish between the two models. 
We will assess detector sensitivity using the $\sigma$ confidence level ($Z$ significance in Ref. \cite{Cowan_2011}), where here $\sigma$ is the standard deviation of the pdf for the hypothesis we are testing \cite{Cowan_2011}. We are interested in both the exclusion power (the probability that $\mu_b$ is observed assuming the signal+background model is true) and the discovery power (the probability that $\mu_{sb}$ is observed assuming the background only model is true).
In the event of null results, the detector will exclude some model/combination of $R_m$ and $R_0$ with $n\sigma$ confidence, where
\begin{equation}
    n=\frac{|\mu_{sb}-\mu_b|}{\sigma_{sb}}.
    \label{exclusion_power}
\end{equation}
Here, $\mu_b$ ($\mu_{sb}$) corresponds to the mean modulation observed by the background (signal+background) model. Where we want to plot the 90\% CL in the DM mass - cross section plane, as is standard for model dependent sensitivity plots, we compute the one sided p-value associated with $n$,
\begin{equation}
    p = \frac{1}{2}\left(1-\text{erf}\left(\frac{n}{\sqrt{2}}\right)\right),
\end{equation}
and require that $p\leq 0.1$.\\
If instead we are interested in the discovery power rather than the exclusion the $\sigma$ confidence level will depend on the uncertainty of the background only distribution, and be given by
\begin{equation}
    n=\frac{|\mu_{sb}-\mu_b|}{\sigma_{b}}.
    \label{discovery_power}
\end{equation}
Thus, the tuneable parameters of a detector are the total mass, live time, detector background, $R_m$, and $R_0$. Typical model dependent analyses will set $R_b$, live time, and mass based on the detector in question, and vary $R_m$ and $R_0$ by setting different values of DM mass ($m_{\chi}$) and cross section ($\sigma_{p}$). However, it is also possible to set $R_m$ and $R_0$, (based on some observation or upper limit, for example) and vary the detector parameters to find the required live time, mass, and background for a detector to observe this particular signal, rather than a specific DM model.
\subsection{Assumptions}
\label{sec:assump}
For both the model dependent and independent sensitivity calculations, we make a number of assumptions for ease of computation and comparison across different detectors. They are compiled in this section for reference. In all cases, we consider three detectors, with the exposure mass, backgrounds and efficiency reported by ANAIS and COSINE, and those planned for SABRE, given in Table \ref{tab:det_ass} with those reported by DAMA for comparison.

\begin{table}[!h]
\caption{Detector properties.}
\label{tab:det_ass}
\centering
\begin{tabular}{@{}llll@{}}
\toprule
Experiment & Mass (kg) & $R_b$ (dru) & $\epsilon$ \\ \midrule
ANAIS  & 112              & 3.2 \cite{ANAIS2018} & Ref. \cite{amare2021} \\
COSINE& 57.5               & 2.7 \cite{Adhikari_2019} & Ref. \cite{adhikari2020eff}  \\
DAMA  & 250              & 0.8 \cite{BERNABEI2020} & Ref. \cite{Bernabei2008} \\
SABRE      & 50               & 0.36 \cite{SABRE_MC} & Ref. \cite{Bernabei2008}\\  \bottomrule
\end{tabular}
\end{table}

\subsubsection{Model dependent assumptions}
The expression we use for the DM interaction rate as a function of nuclear recoil off an elemental target with density (atoms per kg of target material) $N_T$ is
\begin{equation}
    \frac{dR}{dE_R} = N_T\frac{\rho}{m_{\chi}} \int v f_{\rm lab}(\vec{v})\frac{d\sigma_T}{dE_R}d^3v,
\label{eq:general_rate}    
\end{equation}
where the dependence on model choice comes from the selection of couplings constants $\hat{c}_i^{(a)}\hat{c}_j^{(b)}$ in
\begin{equation}
\begin{split}
    \frac{d\sigma_T}{dE_R} 
        &= \frac{m_T}{2v^2}\frac{\sigma_p}{\mu^2_{N}}
            \left[\sum_{i,j}\sum_{N,N'=n,p} \hat{c}_i^{(a)}\hat{c}_j^{(b)}F^{(N,N')}_{ij}(v,q)\right].\\
\end{split}    
\label{idm_rate_full}  
\end{equation}
Note that here $\sigma_p$ is the proton-DM cross section, and the subscript $T$ corresponds to a particular atomic target such as Na \textit{or} I, and $N$ to an individual nucleon. So $\mu_N$ for example is the reduced mass for the nucleon-DM system:
\begin{equation}
    \mu_N = \frac{m_{\chi}m_N}{(m_{\chi}+m_N)}.
\end{equation}
For the simple spin independent model we use here, we take all couplings as non-zero, save for $\hat{c}_1^{(n)}=\hat{c}_1^{(p)}=1$, while the form factors used are taken from the appendix of Ref.\cite{Fitzpatrick_2013}. Under these assumptions, the differential cross section simplifies to
\begin{equation}
\begin{split}
    \frac{d\sigma_T}{dE_R} 
    &=\frac{m_T}{2v^2}\frac{\sigma_p}{\mu^2_{N}}\left(F^{(p,p)}_{11}(q)+F^{(n,n)}_{11}(q)+2F^{(n,p)}_{11}(q)\right).
\end{split}    
\end{equation}
The expressions for these nuclear form factors for $^{23}$Na and $^{127}$I are given in Appendix \ref{app}.\\
We also assume the standard halo model (SHM) for the velocity distribution,
\begin{equation}
    f_{lab}(v)=\frac{1}{(\pi v_0^2)^{3/2}}\exp\left[-\frac{1}{v_0^2}\left(\boldsymbol{v}-\boldsymbol{v}_E\right)^2\right],
\end{equation}
where $v_0=220$ km s$^{-1}$, $v_{\rm esc}=550$ km s$^{-1}$, and $\rho=0.3$ GeV cm$^{-3}$. The Earth's velocity $\boldsymbol{v}_E$ is given by $\boldsymbol{v}_E = \boldsymbol{v}_{\odot}+\boldsymbol{v}_t$, where
\begin{equation}
\begin{split}
    \vec{v}_{\odot} &= v_{\odot}(0,0,1),\\
    \vec{v}_t &= v_t(\sin 2\pi t,\sin\gamma\cos 2\pi t, \cos \gamma\cos 2\pi t),\\
    v_{\odot} &= 232 \text{ kms}^{-1},\\
    v_t &= 30 \text{ kms}^{-1},\\
    \gamma &= \pi/3 \text{ rad}.
\end{split}    
\end{equation}
Here $t$ is a simplification of $(t-t_0)/T$, where for the SHM $t_0=152$ days is the offset from the beginning of the year, and $T$ corresponds to a period of 1 year.\\
These assumptions, when substituted into Eq. \ref{eq:general_rate}, give an expression for the interaction rate as a function of the recoil energy of a target nucleus. In order to transform this into the rate actually observed by the full detector set up, a number of detector specific response functions must be accounted for. These will now be described in detail.\\

For NaI(Tl) scintillators, nuclear recoils experience a quenching relative to the electron recoil events typically used for calibration \cite{bignell2021}. To account for this, a unit change occurs from the keV$_{\rm NR}$ of the nuclear recoil ($E_{R}$) to the electron equivalent energy (keV, or $E_{ee}$), which is what is observed by the detector. The two are related by the quenching factor of the target $Q_T$ via
\begin{equation}
    E_{ee}=Q_T E_R,
\end{equation}
meaning the observable rate in units of keVee is given by
\begin{equation}
    \frac{dR}{dE_{ee}}= \frac{dR}{dE_R}\frac{dE_R}{dE_{ee}}.
\end{equation}

Here we assume (for simplicity) that all three detectors of interest have the same quenching factors as DAMA ($Q_{Na}=0.3$ and $Q_I=0.09$), though we note that deviations from these values that favour an energy dependent quenching factor are suggested by the literature \cite{bignell2021,joo2019,stiegler2017,Xu2015}. We discuss how this might influence the observable rate in Sec. \ref{ssec:qf}.

Another detector effect that will change the observed signal is the finite energy resolution of the detector, a measure of how well it can resolve fine details in the event energy spectrum. As this is related to the observable energy, all units are keV. It is typically based on the FWHM of a specific peak or peaks measured during calibration, and given by a function of the form
\begin{equation}
    \frac{\sigma_E}{E}=\frac{\alpha}{\sqrt{E}}+\beta,
\end{equation}
where $\sigma_E$ is related to the FWHM via FWHM$=2.35 \sigma_E$. We again use the reported values for DAMA given in Ref.\cite{Bernabei2008}:
\begin{equation}
   \sigma_E = \left(0.0091\frac{E_{ee}}{\keV}+0.488\sqrt{\frac{E_{ee}}{\keV}}\right)\keV.
\end{equation}
This is accounted for computationally by smearing the interaction rate with a gaussian distribution, effectively assuming each deposition in the detector at $E_{ee}$ is observed as a spectrum. Thus, the observed rate as a function of $E'$ keV$_{\rm ee}$ (we use $E'$ to distinguish this differential spectrum form from the nuclear recoil $E_R$ and the electron equivalent energy $E_{ee}$ being integrated over) becomes
\begin{equation}
    \frac{dR}{dE'}=\frac{1}{\sqrt{2\pi}}\int_0^{\infty}\frac{1}{\sigma_E}\frac{dR}{dE_{ee}}\exp{\left[\frac{-(E'-E_{ee})^2}{2(\sigma_E)^2}\right]} dE_{ee}.
\label{eq:full_res}
\end{equation}

Finally, we account for the detector's thresholds and efficiency in Eqs \ref{back_n} and \ref{sigback_n}. For DAMA this is given in Ref. \cite{Bernabei2008}:
\begin{equation}
    \epsilon(E)=\left\{\begin{matrix}
0.0429 E + 0.657 & \text{for $E<8$ keV}\\  
1 & \text{otherwise}
\end{matrix}\right.\\
.\end{equation}
The values assumed for COSINE and ANAIS are given in Refs. \cite{amare2021,adhikari2020eff}, while for SABRE we assume DAMA efficiency. For the cases we consider, we look only at the 2-6 keV region to be consistent across all the detectors, as this is the only reported value for SABRE's background.\\
Thus, the observable rate at as a function of energy $E'$ at a given detector is
\begin{equation}
    \frac{dR}{dE'}=\frac{\epsilon(E')}{Q_T}\frac{1}{(2\pi)^{1/2}}\int_{0}^{\infty}\frac{dE_{ee}}{\sigma_{E}}\frac{dR}{dE_{R}}\exp\left[\frac{-(E'-E_{ee})^2}{2(\sigma_{E})^2}\right],
\end{equation}
and the total observed rate in a finite energy bin of width $\Delta E = 2\delta E$ (in cpd/kg/keV)  is
\begin{equation}
\begin{split} 
&R(E) = \frac{1}{\Delta E}\int_{E-\delta E}^{E+\delta E}\frac{dR}{dE'} dE'.\\
         &
\end{split}
\label{ob_diff}    
\end{equation}
This can be split into the constant and modulating components by projecting onto a cosine function with the expected period and offset for DM.\\
Note that these expressions have been defined assuming a single element target $T$. In the case of a composite target such as NaI, the total rate is given by adding the rates from each individual target weighted by their contributing mass fraction:
\begin{equation}
    \frac{dR_\text{Tot}}{dE'}=\sum_{i}\frac{m_i}{m_\text{Tot}}\frac{dR_i}{dE'}, 
\end{equation}
where here $i$ is Na or I for the detectors of interest for this study.

\subsubsection{Model independent assumptions}
\label{sssec:indep_ass}
For the model independent analysis, we assume that all detectors should observe exactly the same rate as reported by DAMA as they use the same target and a similar setup. This allows us to conduct analysis without needing to assume some particular DM model, and just assess how easily modulation of a certain magnitude can be observed over a constant background signal, regardless of its origin.\\  
In that case we we can take the signal rate to be $R(t) = S_0+ S_m\cos{\omega t}$ for the reported bins, where $S_m$ is given explicitly in Ref. \cite{Bernabei2018}. For the detectors currently operating (ANAIS and COSINE), the constant rate reported is, in reality, a combination of the background and any constant DM rate if it exists. As such, for the model independent study we take the values reported in Table \ref{tab:det_ass} to be $R_b+S_0$, and test how well these experiments can constrain a modulation given their observed constant rate.\\
However, SABRE is yet to start taking data, and the background value in Table \ref{tab:det_ass} is based on simulation rather than observation, and so does not account for any existing additional constant rate. To account for this, we use as a constant rate the upper limits in this region reported by DAMA in Ref. \cite{BERNABEI2020}.
The relevant rates are given in Table \ref{tab:dama_rates}.

\begin{table}[!h]
\caption{Modulating \cite{Bernabei2018} and upper limits on constant \cite{BERNABEI2020} DAMA rates.}
\label{tab:dama_rates}
\centering
\begin{tabular}{@{}ccc@{}}
\toprule
~Energy (keV) & $S_m$ (cpd/kg/keV) & $S_0$ (cpd/kg/keV)~ \\ \midrule
1.0 - 1.5      & 0.0232 $\pm$ 0.0052  & 0.80              \\
1.5 - 2.0      & 0.0164 $\pm$ 0.0043  & 0.80             \\
2.0 - 2.5      & 0.0178 $\pm$ 0.0028  & 0.24              \\
2.5 - 3.0      & 0.0190 $\pm$ 0.0029  & 0.24             \\
3.0 - 3.5      & 0.0178 $\pm$ 0.0028  & 0.12              \\
3.5 - 4.0      & 0.0109 $\pm$ 0.0025  & 0.12              \\
4.0 - 4.5      & 0.0110 $\pm$ 0.0022  & 0.12             \\
4.5 - 5.0      & 0.0040 $\pm$ 0.0020  & 0.12              \\
5.0 - 5.5      & 0.0065 $\pm$ 0.0020  & 0.12              \\
5.5 - 6.0      & 0.0066 $\pm$ 0.0019  & 0.12             \\ \bottomrule
\end{tabular}
\end{table}

\section{Results}
\label{sec:results}
Our sensitivity results are split into two sections, model dependent and independent. The former are the more traditional presentation, selecting a model (here, the `standard' spin independent elastic WIMP) and varying $m_{\chi}$ and $\sigma_p$ - equivalent to changing $R_m$ and $R_0$. The latter demonstrate the exclusion power of various NaI experiments as a function of their live time, setting $R_m$ and $R_0$ to the values reported by DAMA.
\subsection{Model dependent}
\label{sec:moddep}

The choice of a particular model will dictate the observable $R_m$ and $R_0$ for some combination of DM mass and cross section, $m_{\chi}$ and $\sigma_{p}$. To produce the usual sensitivity curves, we loop through a set range of $m_{\chi}$ and $\sigma_{p}$, and compute the p-values of $\mu_b$ (the mean of the background only distribution) in the signal+background distribution. A detector will be sensitive with 90\% confidence to the combinations that produce $P(x=\mu_b)\leq 0.1$.

\begin{figure}[!h]
    \centering
    \includegraphics[width=0.5\textwidth]{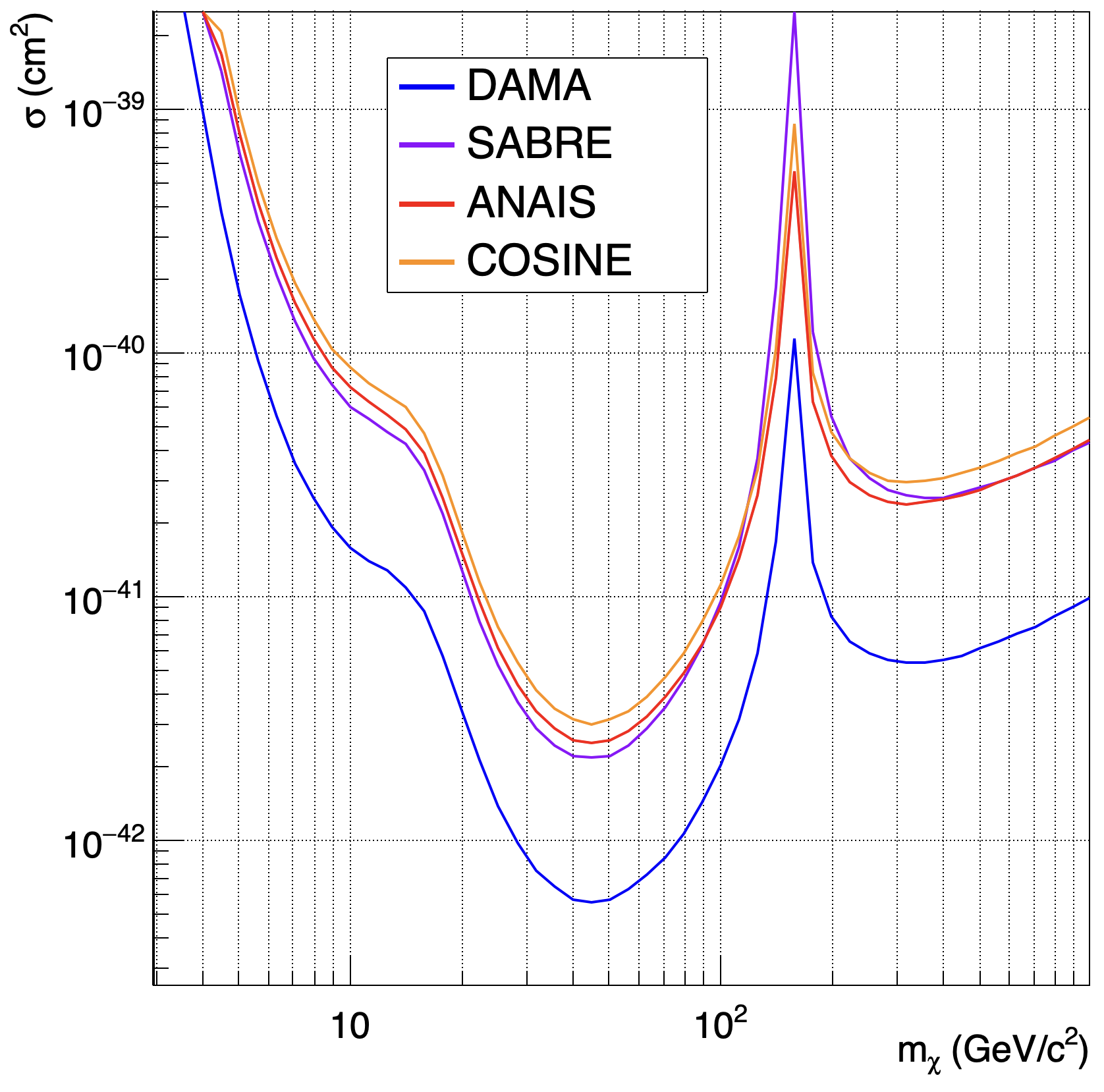}
    \caption{90  \% C.L. modulation sensitivity of various NaI detectors by 2025, assuming all three are continuously taking data. We assume a spin independent elastic WIMP. For each detector the exposure mass and $R_b$ are shown in Table \ref{tab:det_ass}. The total live time assumed for each detector by 2025 is 18 yrs (DAMA), 9 yrs (COSINE), 8 yrs (ANAIS), and 3 yrs (SABRE).}
    \label{fig:model_dep_sens}
\end{figure}

This method is applied to DAMA, ANAIS, COSINE, and SABRE to project their sensitivity by 2025 (assuming all three detectors are continuously taking data until then). The results for this are shown in Fig. \ref{fig:model_dep_sens} for the standard spin independent DM model. This demonstrates that although all the detectors have the same target, there are still combinations of DM parameter space that will produce a statistically significant modulation in some detectors, but not in others despite (in some cases) a longer live time.

\begin{figure}[!h]
    \centering
    \includegraphics[width=0.5\textwidth]{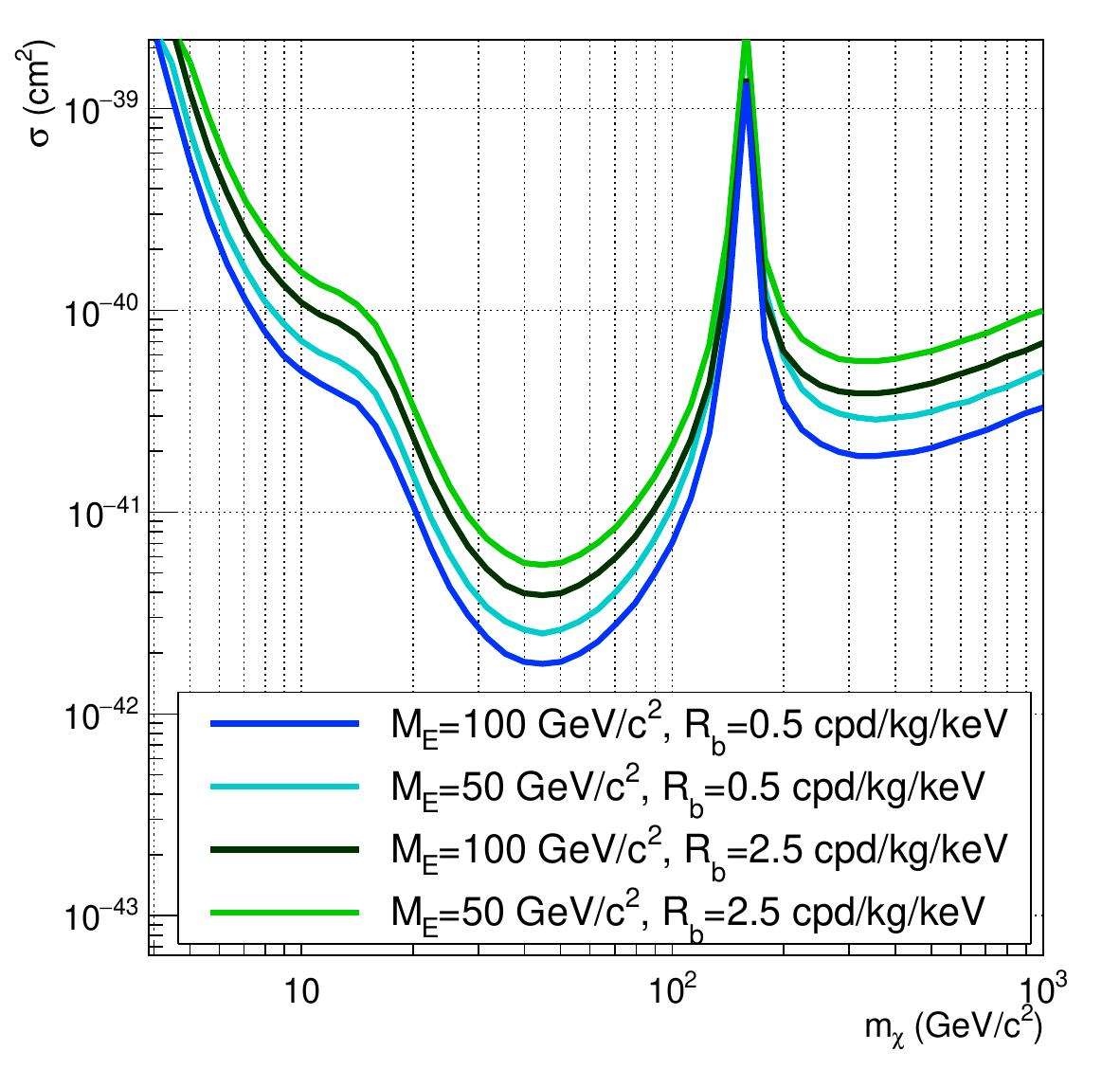}
    \caption{90  \% C.L. modulation sensitivity of different NaI mass and background combinations for 3 years live time.}
    \label{fig:model_dep_cons}
\end{figure}

In order to decouple the relative live times, we also compare the effect of changing just the mass and background levels for a NaI target in Fig. \ref{fig:model_dep_cons}. In doing so, we observe the standard scaling with $\sqrt{R_b/M_E}$ is recovered, validating this computation method.

\begin{figure}[!h]
    \centering
    \includegraphics[width=0.5\textwidth]{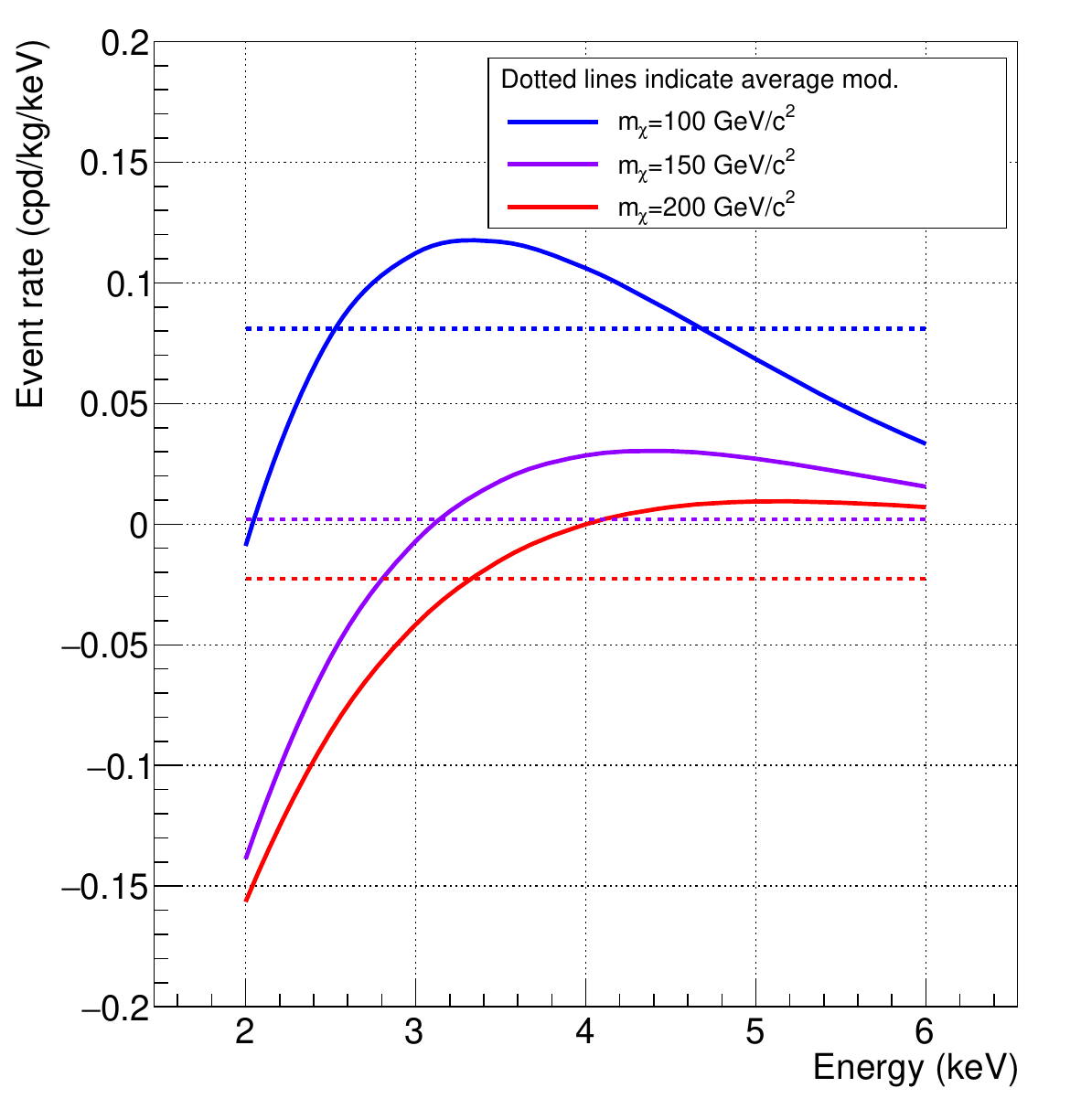}
    \caption{Modulation rate both as a spectrum and averaged over the region of interest for various DM masses and $\sigma_p=2.5^{-40}$ cm$^2$.}
    \label{fig:spike}
\end{figure}

Note that the peak occurring at $m_{\chi}\sim 150$ GeV/c$^2$ in both Figs. \ref{fig:model_dep_sens} and \ref{fig:model_dep_cons} is due to our assumption of a single energy bin from 2-6 keV, as we have assumed a flat background hypothesis for simplicity. At this mass range the net modulation in our energy region of interest switches from positive to negative (see Fig. \ref{fig:spike}), and so averaging over the energy bins erroneously gives a modulation amplitude close to zero, drastically reducing the sensitivity. More proper analysis requires knowledge of the background model as a function of energy not publicly available.

The exposure mass and detector backgrounds assumed for this analysis are given in Table \ref{tab:det_ass}. Despite having the lowest target mass, SABRE has the greatest sensitivity of the three new detectors to the annual modulation of these models across the full parameter space. This is due its low background, which is projected to be approximately an order of magnitude lower than that of ANAIS, the next most sensitive detector, which gives it the most competitive $\sqrt{R_b/M_E}$ value. This demonstrates the importance of considering the ratio of mass to background when increasing the active mass of a detector, as adding mass that has a high background can produce a detector with a larger value of $\sqrt{R_b/M_E}$, meaning the detector will take longer to produce significant results than it would with the smaller mass. This can also be seen in the decision made by COSINE to reduce their active mass by excluding a number of larger mass, higher background crystals \cite{Adhikari_2019}.

The analysis described in this section is typical in the production of exclusion plots, and is the only way to compare detectors using different targets. The downside of this method is that it remains possible to tailor increasingly complex DM models to explain the lack of a modulating signal at detectors other than DAMA. To properly exclude (or support) this observation, we need to understand and compare the exclusion power of other NaI detectors for the DAMA rate, in the event they return null results.

\subsection{Model independent}
\label{ssec:modindep}
\begin{figure}[!h]
    \centering
    \includegraphics[width=0.5\textwidth]{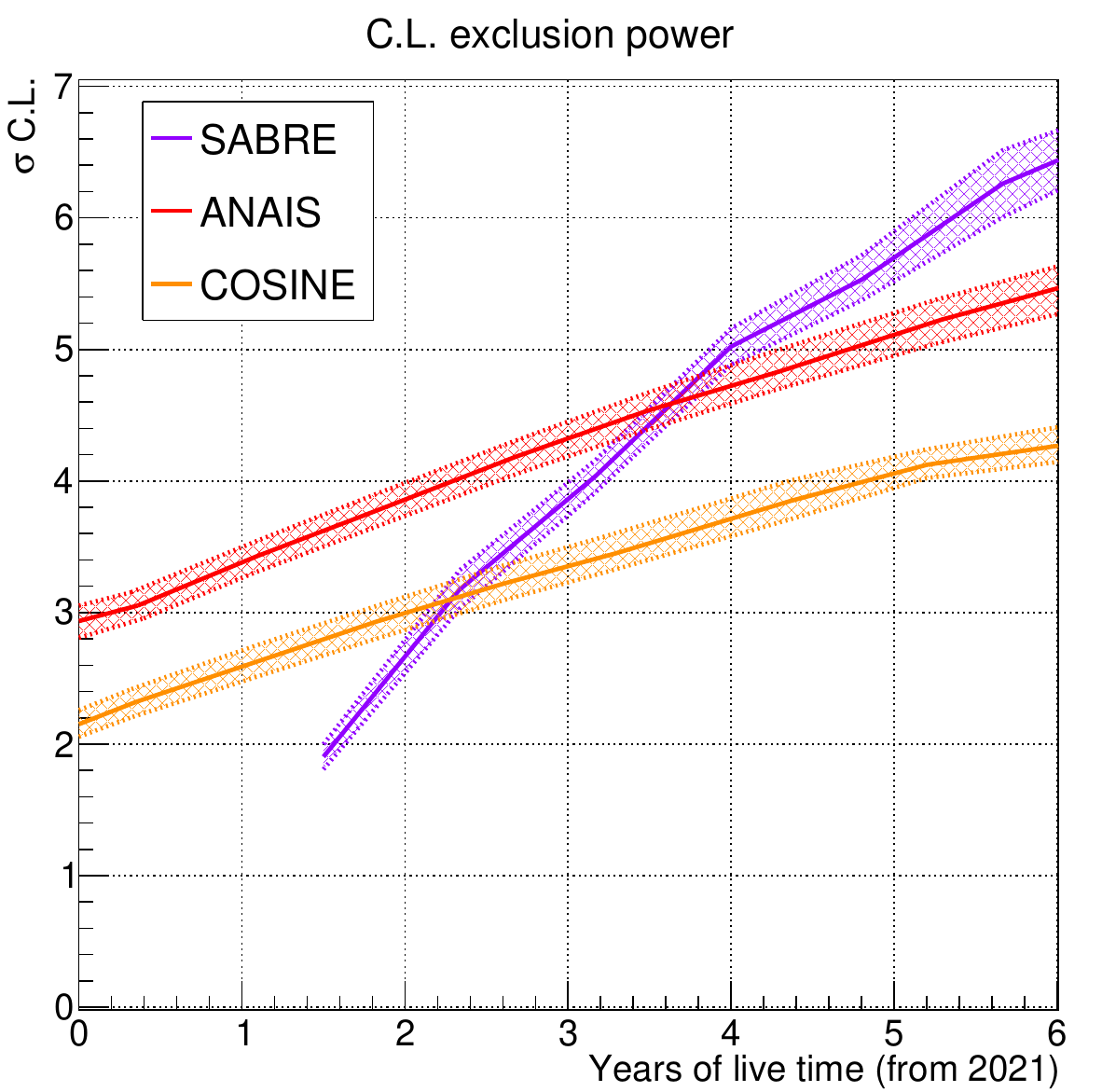}
    \caption{Conservative exclusion power of NaI detectors, as a function of live time. Cross hatched regions indicated 1$\sigma$ statistical uncertainty bands.}
    \label{fig:model_indep_excl}
\end{figure}

\begin{figure}[!h]
    \centering
    \includegraphics[width=0.5\textwidth]{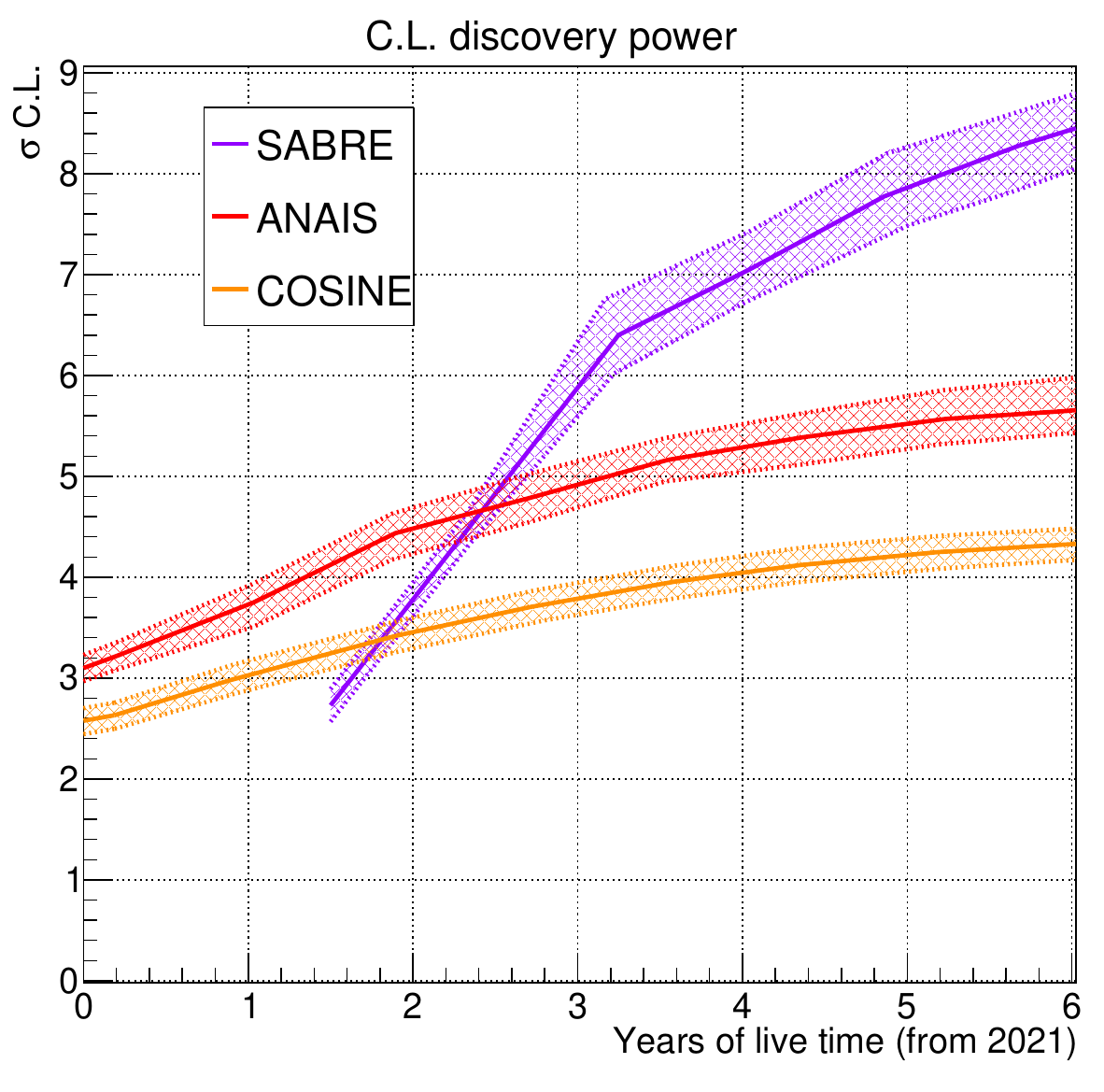}
    \caption{Conservative discovery power of NaI detectors, as a function of live time. Cross hatched regions indicated 1$\sigma$ statistical uncertainty bands.}
    \label{fig:model_indep_disc}
\end{figure}

Detectors that use the same target should theoretically all observe the same interaction rate. In reality, the efficiency and resolution may cause some variation between detectors, but we can consider the DAMA data a minimum observable rate, as the newer detectors tend to improve upon both values \cite{adhikari2020eff,Amare2019eff}, particularly above 2 keV. We can take advantage of this, and instead of assuming some model, DM mass, and cross section to construct $R_m$ and $R_0$, we use the DAMA values given in Ref.\cite{Bernabei2018}, which should also be visible at other NaI detectors. We also do not need to adopt any specific values to account for the detector effects, provided we assume that they are consistent with DAMA's\footnote{Provided they are roughly the same across the detectors, the actual value of effects like quenching factor or resolution will influence the \textit{interpretation} of the result (i.e., attributing it to a particular DM mass), not the magnitude of the observable rate itself. See Sec. \ref{ssec:qf} for a discussion of how this might change if the detector effects do differ between experiments.}. Using this, we can find the $n\sigma$ confidence level exclusion or discovery power of the DAMA rate after a given live time.\\ 
The exclusion power of the various NaI detectors is shown in Fig. \ref{fig:model_indep_excl}, discovery power in Fig. \ref{fig:model_indep_disc}, and the properties assumed for each of detector are given in Table \ref{tab:det_ass}. Key and present sensitivity milestones are also shown in Table \ref{tab:detector_prop}. These results again assume a single energy bin from 2-6 keV, and a constant background over that region as reported in the provided references. We also note that the present and predicted exclusion power we find for both ANAIS and COSINE are consistent with those previously presented by both collaborations \cite{amare2021,Adhikari_2019}. The 1$\sigma$ uncertainty is found by repeating the MC simulation and calculation of significance a number of times, and computing the standard error of the mean (taking this as $\sigma$).

\begin{table}[!h]
\caption{Present exclusion C.L. for the detectors, and years for significant exclusion (Fig. \ref{fig:model_indep_excl}) and discovery C.L. (Fig. \ref{fig:model_indep_disc}).}
\label{tab:detector_prop}
\centering
\begin{tabular}{@{}llll@{}}
\toprule
~Exp. & Present excl. & For 3$\sigma$ excl. & For 5$\sigma$ disc.\\ \midrule
~ANAIS  &  2.9$\sigma$  & $\sim$ 3 yrs & $\sim$ 7 yrs \\
~COSINE &  2.2$\sigma$  & $\sim$ 5 yrs & $>$ 7 yrs  \\
~SABRE      &0            & $\sim$ 2 yrs& $\sim$ 2 yrs\\  \bottomrule
\end{tabular}
\end{table}

For this model independent analysis, we see the same trend as in Fig. \ref{fig:model_dep_sens}, where the low background of SABRE makes it the most sensitive experiment, despite also having the lowest target mass. This is even more obvious considering the exclusion vs. discovery plots. As explained in Sec \ref{sec:sensecalc} in Eq. \ref{exclusion_power} and \ref{discovery_power}, the former depends on the uncertainty in \textit{signal+background}, while the latter relies only on \textit{background} uncertainty. As such, the improvements from lowering the background are much clearer in the case of discovery power, which explains why the SABRE experiment is able to ramp up much faster than COSINE and ANAIS.\\
Focusing on the discovery plot in Fig. \ref{fig:model_indep_disc}, this gives the significance of a modulating signal compatible with DAMA's, should it be observed by any of the detectors. Clearly, a lower background gives a significant advantage, in terms of time the experiment needs for statistically significant results, as well as the laboratory space it will ultimately take up.\\
As noted in Sec. \ref{sssec:indep_ass}, these limits from SABRE are conservative, as the simulated background spectrum requires the injection of an additional constant component nominally attributable to DM.

\subsubsection{Time dependent backgrounds}
\begin{figure}[!h]
    \centering
    \includegraphics[width=0.45\textwidth]{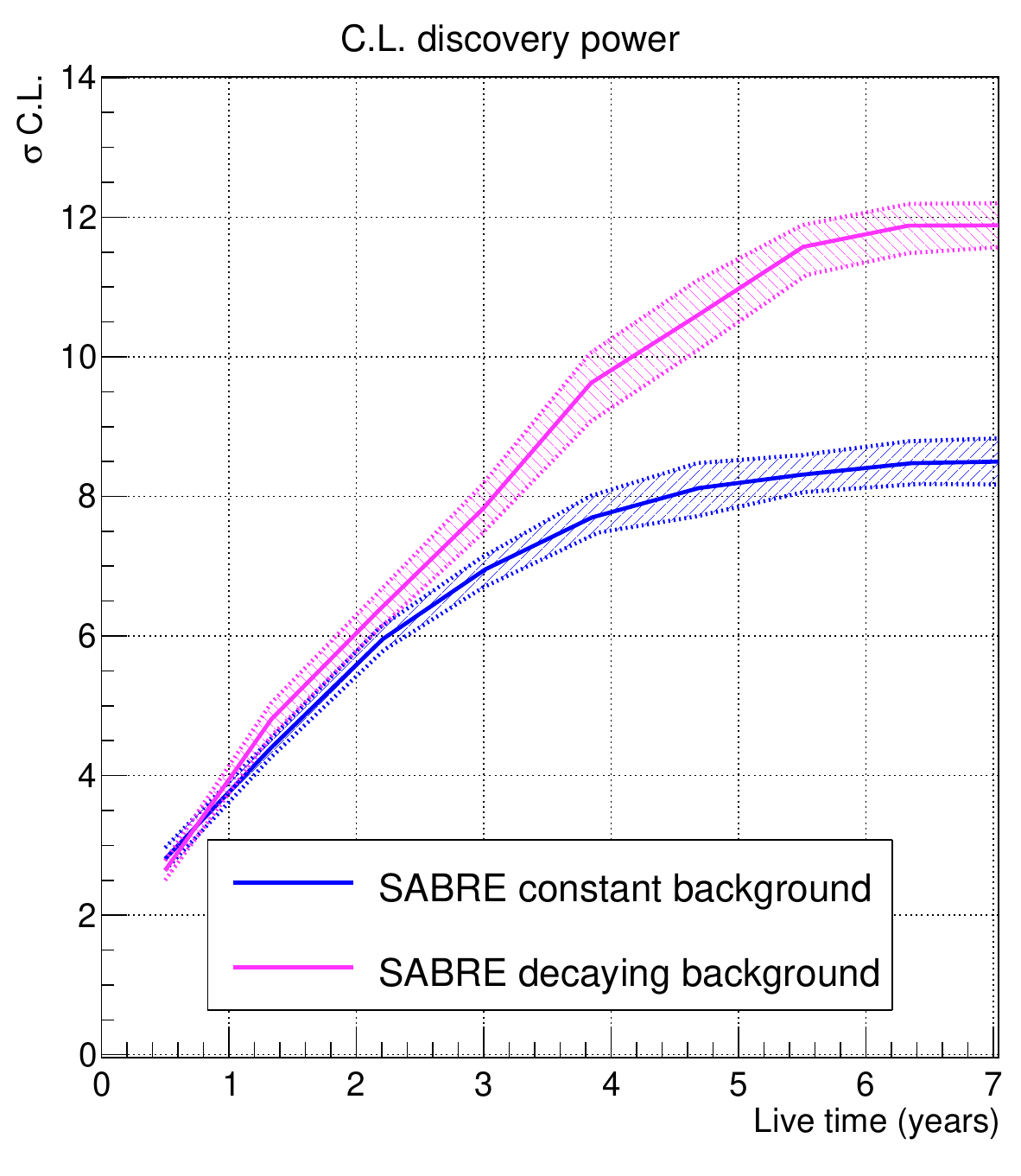}
    \caption{Discovery power of SABRE for a constant and decaying background modelled by Eq. \ref{eq:decay}.}
    \label{fig:const_v_decay}
\end{figure}

The presentation and methodology of the model independent results makes it a convenient way to compare constant and time dependent backgrounds over a detectors lifetime. Due to cosmic activation of NaI(Tl) crystals during travel, they are likely to contain tritium, producing an exponentially decaying background similar to that observed by ANAIS in Ref. \cite{amare2021}. We demonstrate how the consideration of a decaying background might change the sensitivity of an experiment by considering two background models for SABRE - the constant value of 0.36 cpd/kg/keV considered in preceding sections, and a time dependent one given by
\begin{equation}
    R_b(t) = 0.36(0.5+0.5e^{-t/\tau}),
\label{eq:decay}    
\end{equation}
where $\tau$ is the half life of tritium (4503 days), which contributes half of the background at the start of the experiment. The influence of this on the discovery power of SABRE is shown in Fig. \ref{fig:const_v_decay}. The decaying background gives an improvement compared to the constant case, particularly over longer time scales. This clearly demonstrates the necessity of careful background measurement and modelling to produce accurate exclusion and discovery limits for an experiment.

\subsection{Mass and background requirements}
\label{sec:timely}
The model independent sensitivity method can also be used to compute the mass and background required for a detector to exclude the DAMA results with a given confidence level within a specified period of time. For example, assuming that we want an experiment to have exclusion power of 3$\sigma$ within three years, we can plot the maximum allowable background that will give this sensitivity as a function of exposure mass. Alternatively, it can be used to find the minimum exposure mass in the event we cannot get below some particular background threshold.\\

\begin{figure}[!h]
    \centering
    \includegraphics[width=0.48\textwidth]{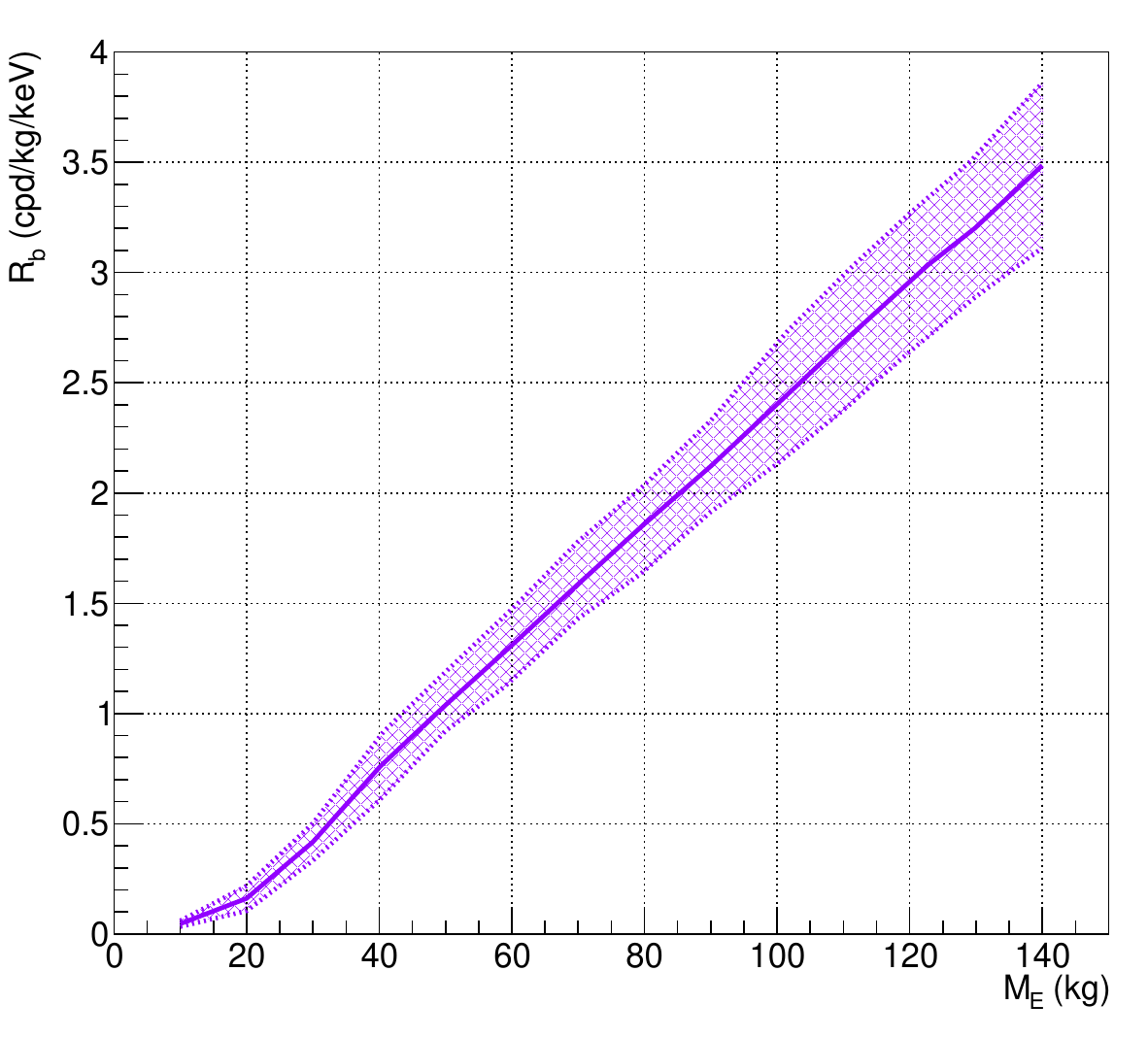}
    \caption{Maximum background and minimum mass combinations that provide 3$\sigma$ exclusion of DAMA within three years of operation. Hatched bands give 1$\sigma$ uncertainty. If a detector's properties fall below the line, it will have the requisite sensitivity.}
    \label{fig:max_rb}
\end{figure}

These results are given in Fig. \ref{fig:max_rb}. If a collaboration has only enough time/resources to grow $\sim$ 65 kg of crystal, they will require an experiment background of below 1.5 cpd/kg/keV in order to have 3 $\sigma$ exclusion power of the the DAMA signal within three years. Conversely, if they know they are able to achieve a background of 0.5 cpd/kg/keV, they need only about 30 kg of crystal for the same time and sensitivity requirements.\\
Calculations and plots of this kind may prove useful at the R\&D or upgrade stages of NaI detectors to inform or validate choices made surrounding target masses and background reduction.

\subsection{Quenching factor influence}
\label{ssec:qf}
\begin{figure}[!h]
    \centering
    \includegraphics[width=0.5\textwidth]{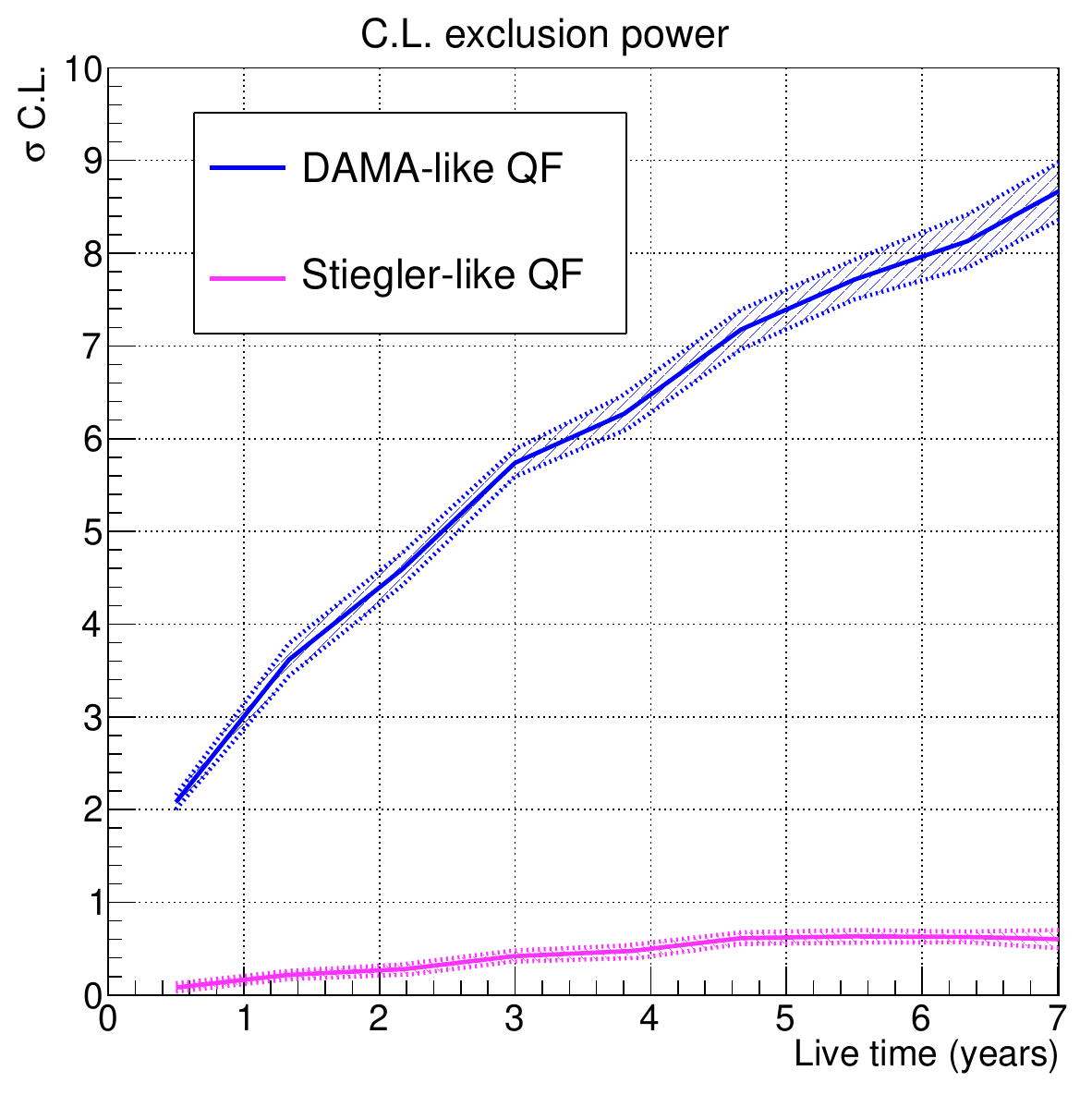}
    \caption{Influence of a crystal dependent quenching factor for an extreme, illustrative DM model with $m_{\chi}=10$ GeV/c$^2$ and $\sigma_{p}=2.9\times 10^{-40}$}
    \label{fig:qf_simple}
\end{figure}

As the quenching factor acts as a conversion between recoil and observed energy, clearly it's value will influence the signal detectable by an experiment. Measurements of the Na quenching factor over recent years \cite{bignell2021,joo2019,stiegler2017,Xu2015} have demonstrated an energy dependence of this factor, though not all the measurements are consistent with each other, suggesting that quenching factor may be a crystal dependent quality \cite{BERNABEI2020,amare2021}. If this is the case, exactly the same DM particle will produce a different observable signal in different detectors. The exact difference will depend strongly on the exact model and mass of the DM, but preliminary studies have demonstrated that in some cases a crystal with a QF similar to that observed by Stiegler et al. in Ref. \cite{stiegler2017} will see a modulation an order of magnitude smaller than that observed by a crystal with $Q_{\rm Na}=0.3$. For example, values of $m_{\chi}=10$ GeV/c$^2$ and $\sigma_{p}=2.9\times 10^{-40}$ cm$^2$ for a spin independent WIMP produce a modulation of 0.0118 cpd/kg/keV assuming the DAMA QF, but 8.64$\times 10^{-4}$ cpd/kg/keV for the QF model presented in Ref. \cite{stiegler2017}.
We demonstrate how SABRE's exclusion power would change for these two QFs in this case in Fig. \ref{fig:qf_simple}. \\

If indeed this is a crystal dependent factor as has been suggested, a truly model independent test of DAMA becomes impossible. One would need to know the QFs of all NaI crystals involved (DAMA's as well as the new experiment) and select some model to understand how DAMA's observed $\sim$0.01 cpd/kg/keV modulation might manifest in this new set up. Under this paradigm, the findings presented in Sec. \ref{ssec:modindep} can be viewed as the maximum possible exclusion or discovery limits. In reality, certain mass and model combinations will produce weaker constraints, but will never be able to produce any stronger, unless the QF is greater than that reported by DAMA in Ref. \cite{Bernabei:1996} (a larger quenching factor corresponds to the ability to probe lower recoil energies). Further study is ongoing to understand the crystal dependence of quenching factor \cite{cintas2021}, and how this might influence the ability of null results to exclude DAMA's observation. The results presented here are initial, illustrative findings that may be of interest given the topic of this paper.

\section{Conclusions}
\label{sec:conclusions}
We have presented here the comparative sensitivity of NaI detectors computed in both a model dependent and independent way, and illustrated the basic influence different background models and quenching factors might have on these results. In both cases, for both exclusion and discovery limits the lowest background experiment (SABRE) has performed the best of the three experiments, despite being the one with the lowest exposure mass. This is due to the dependence of sensitivity on the ratio of $R_b/M_E$ rather than just the exposure mass or background. Thus, while scale up of these NaI based detectors is important,  a lower background must also be pursued in order to observe the small modulation in an already low interaction rate, further motivating the purification and veto techniques presently explored by these collaborations.
In addition, the Monte Carlo method used to generate expected observations for different modulation signals can be used to model more complex and realistic backgrounds such as the specific time dependence of the individual crystal detectors, where such information is available.\\
Finally, this study demonstrates that should the projected exposure mass and backgrounds be achieved, SABRE will be positioned to provide statistically significant exclusion or discovery of the DAMA signal within 2-3 years of commenced data taking.
In this event (and even more so in the event of a positive DM-like signal), it will be perhaps beneficial and interesting to compare the results of the Northern and Southern hemisphere experiments and data, to further elucidate clues as to nature of the modulating DAMA signal - DM or not.

\section*{Acknowledgments}
 The authors would like to thank Phillip Urquijo for his review of the paper. This work was supported by the ARC through grant CE200100008. MJZ is also supported through the Australian Government Research Training Program Scholarship. Both authors are members of the SABRE collaboration.

\appendix
\section{Nuclear form factors}
\label{app}
These form factors are given in terms of the dimensionless $y=(qb/2)^2$, where $b$ is the harmonic oscillator parameter:
\begin{equation}
    b = \sqrt{\frac{41.467}{45A^{-1/3}-25A^{-2/3}}} \text{ fm}.
\end{equation}
\textbf{$^{23}$Na}:
\begin{equation}
\begin{split}
    F^{(p,p)}_{11}(q) &= e^{-2y}\left(120-180y+87y^2-17y^3+1.2y^4\right)\\
    F^{(p,n)}_{11}(q) &= e^{-2y}\left(130-200y+100y^2-20y^3+1.5y^4\right)\\
    F^{(n,n)}_{11}(q) &= e^{-2y}\left(140-220y+120y^2-25y^3+1.8y^4\right)\\
\end{split}    
\end{equation}
\textbf{$^{127}$I}:
\begin{equation}
\begin{split}
    F^{(p,p)}_{11}(q) = &e^{-2y}(2800-10000y+14000y^2-9800y^3\\
    &+3800y^4-840y^5+100y^6-6.3y^7+0.15y^8)\\
    F^{(p,n)}_{11}(q) = &e^{-2y}(3900-15000y+23000y^2-18000y^3\\
    &+7900y^4-2000y^5+290y^6-23y^7+0.75y^8\\
    &-0.0048y^9)\\
    F^{(n,n)}_{11}(q) = &e^{-2y}(5500-23000y+38000y^2-32000y^3\\
    &+16000y^4-4600y^5+790y^6-75y^7+3.3y^8\\
    &-0.041y^9+0.00015y^{10})\\
\end{split}    
\end{equation}

%

\label{Bibliography}

\end{document}